\documentclass[usegraphicx]{mn2e}

\usepackage{times}

\newcommand{\half}{{\textstyle\frac{1}{2}}}

\title[Stellar blending: biases and binarity]{Photometric biases due 
to stellar blending: implications for measuring
distances, constraining binarity and detecting exoplanetary transits
}
\author[L. L. Kiss \& T. R. Bedding]{L. L. Kiss\thanks{E-mail:
l.kiss@physics.usyd.edu.au}\thanks{On leave from University of Szeged, Hungary} \& T. R. Bedding\\
\\
School of Physics, University of Sydney 2006, Australia}

\begin{document}

\date{Accepted ... Received ..; in original form ..}


\maketitle

\begin{abstract}

We investigate blending, binarity and photometric biases in crowded-field CCD 
imaging. For this, we consider random blend losses, which correspond to the total
number of stars left undetected in unresolved blends. We present a simple formula to
estimate blend losses, which can be converted to apparent magnitude biases using the
luminosity function of the analyzed sample. Because of the used assumptions, our results
give lower limits of the total bias and we show that in some cases even these limits
point toward significant limitations in measuring apparent brightnesses of ``standard
candle'' stars, thus distances to nearby galaxies. A special application is presented
for the OGLE-II $BVI$ maps of the Large Magellanic Cloud. We find a
previously neglected systematic bias up to 0\fm2--0\fm3 for faint stars 
($V\sim18\fm0-19\fm0$) in 
the OGLE-II sample, which affects LMC distance measurements using RR~Lyrae and red clump stars.
We also consider the 
effects of intrinsic stellar correlations, i.e. binarity, via calculating two-point
correlation functions for stellar fields around seven recently exploded
classical novae. In two cases, for V1494~Aql and V705~Cas, the reported close optical
companions seem to be physically correlated with the cataclysmic systems. 
Finally, we find significant blend frequencies up to 50--60\% in the samples of
wide-field exoplanetary surveys, which suggests that blending calculations are highly advisable to
be included into the regular reduction procedure. 

\end{abstract}


\section{Introduction}

Classical crowded field photometry attempts to detect and measure brightnesses
of individual stars that are heavily affected by the presence of close
neighbours. For ground-based observations, crowding depends on the angular
density of objects and the atmospheric seeing conditions. Stellar blending
(unresolved imaging of overlapping stars) can be a significant component of the
total ambiguity known as the confusion noise  -- see Takeuchi \& Ishii (2004)
for an excellent historic review and a general formulation of the source
confusion statistics. Recent studies for which blending was important include
measuring: the luminosity  function of individually undetectable faint stars
(Snel 1998); the extragalactic Cepheid period-luminosity relation (Mochejska et
al. 2000, Ferrarese et al. 2000, Gibson et al. 2000); and Cepheid light curve
parameters (Antonello 2002). Extensive investigations can also be found about
blending and microlensing  surveys (Alard 1997; Wozniak \& Paczynski 1997; Han
1997, 1998; Alcock et al. 2001).  Stellar blending in general is difficult to
model, because significant contribution may be due to physical companions,
which are common among young stars,  including Cepheids (Harris \& Zaritsky
1999, Mochejska et al. 2000). 

Here we attempt to determine the effects of random blending with a new approach
that includes corrections for the excess of double stars.  This work was
motivated by recent cases in which stellar blending played a degrading role.
For instance, wide-field photometric surveys of the galactic
field are characterized by confusion radii of
10-20$^{\prime\prime}$ (Brown 2003) and can suffer from strong blending, even
in regions far from the  galactic plane. Other examples include the presence of 
a close optical companion of the classical nova
V1494~Aql (at a separation of about 1\farcs5), which heavily affected late
light curves of the eclipsing  system (Kiss et al. 2004). 

Inspired by these problems and the availability of deep, all-sky star
catalogues like the USNO B1.0 (Monet et al. 2003), we decided to perform simple
calculations in terms of observational parameters such as the confusion radius
-- constrained by the seeing or the pixel size of the detector -- and stellar
angular density.  To emphasize the importance of the problem, we present
magnitude and  amplitude biases for unresolved blends of $\Delta m$=0, 1, 2 and
3 mag in  Table \ \ref{biases}. Though fairly trivial, the numbers clearly show
that even for $\Delta m$=3 mag, blending can affect brightness and variability
information to a highly significant extent. Also, these numbers span the
magnitude ranges in which we are interested: usually 3 to 5 mag wide samples
are considered for the chance of random blending. This is a different approach
compared to the case of general blending, which  includes very faint blends,
too. For instance, Han (1997) showed that for a typical 1\farcs5 seeing disk
towards the galactic Bulge, model luminosity functions of the Milky Way predict
$\sim$36 stars within that area; of these, only ~0.75\% is expected to be
brighter than 18 mag. Obviously, those faint blends do not affect photometry of
bright stars. This study focuses on a much more specific problem: for a given
field of view and  confusion radius, what can be derived about random blending
probabilities from the  observed stellar angular distributions? Can we assign
systematic errors  based on these probabilities?  

The paper is organized as follows. In Sect.\ 2 we discuss blending  in random
stellar fields. We present a simple formula to estimate random blend losses, which
has been tested by extensive simulations. We also discuss the effects of visual
double stars by determining the  two-point correlation. As an application, in
Sect.\ 3 we investigate probable photometric biases in deep  OGLE-II $BVI$
observations of the Large Magellanic Cloud (Udalski et al. 2000).  In Sect.\ 4 we
analyse stellar  fields around seven recently erupted classical novae, all located
in densely populated regions near to the galactic plane. We investigate blending
rates in photometric survey programs HAT (Bakos et al. 2002, 2004),
STARE\footnote{See also  {\tt
http://www.hao.ucar.edu/public/research/\\public/stare/stare.html}}  (Brown 2003,
Alonso et al. 2003) and ASAS (Pojmanski 2002) in Sect.\ 5. Concluding remarks are
given in Sect.\ 6.

\begin{table}
\begin{center}
\caption{Biases in apparent magnitude ($m_{\rm obs}$) and amplitude of 
variation ($A_{\rm obs}$) for blending stars of magnitude difference $\Delta m$.}
\label{biases}
\begin{tabular}{lrrrrr}
\hline
     & unblended & $\Delta m=0$ & $\Delta m=1$ & $\Delta m=2$ & $\Delta m=3$ \\
\hline
$m_{\rm obs}$ & 0\fm00 & $-$0\fm75 & $-$0\fm36 & $-$0\fm16 & $-$0\fm07 \\
$A_{\rm obs}$ & 1\fm00 & 0\fm39 & 0\fm61 & 0\fm80 & 0\fm91 \\
$A_{\rm obs}$ & 0\fm10 & 0\fm05 & 0\fm07 & 0\fm09 & $\sim$0\fm10\\
$A_{\rm obs}$ & 0\fm01 & 0\fm005 & 0\fm007 & 0\fm009 & $\sim$0\fm01 \\
\hline
\end{tabular}
\end{center}
\end{table}

\section{Random blending and binarity: basic relations}

Hereafter we distinguish two cases that need different approaches:

\medskip

\noindent {\bf Case 1:} There is only one dataset to analyse for blending
probability, characterized by a typical confusion radius; no additional
information exists based on a catalogue or high-resolution imaging with much
smaller confusion radius.

\medskip
 
\noindent {\bf Case 2:} We can compare the observations with an additional source
of data, which can be considered as unbiased by random blending.

\medskip

\noindent Case 1 refers to those studies where the completeness of a catalogue 
is investigated or where the observations were deeper than any existing
catalogue. Examples discussed in this paper include the deep OGLE-II $BVI$ map
of the Large Magellanic Cloud and galactic novae in the USNO  B1.0 catalogue.
Case 2 corresponds to the exoplanetary surveys with very large confusion radii
(up to 20$^{\prime\prime}$), which can be well characterized using whole-sky
star catalogues of a magnitude better angular resolution. Case 2 is much
simpler because one can always check whether an interesting object is a blend
or not. However, if ensemble properties of large groups of stars are
considered, blending must be taken into account because high fractions of stars
are in fact blends when observed at large confusion radius. 

\subsection{Random blend loss}

The idea behind our calculations is the following. When taking an image of a
very crowded field, the detection efficiency is limited by the confusion radius
($r_c$), which is the smallest angular distance between two resolvable stars.
If the distance between two neighbours is smaller than $r_c$, we detect only
one object. This means the number of detected stars, $N_d$, will be smaller
than $N$ by the number of objects in an area $\delta S=\pi r_c^2$.  The
difference $N-N_d$ is the blend loss. Because of this, the detected ``stars''
will appear to be brighter than they really are. Our aim is to estimate the
number of stars lost due to random merging and the corresponding mean magnitude
biases.

Let us consider a sample of $N$ single stars spread randomly over a field  
with area $S$. The mean number of stars within an area $\delta S$ is

\begin{equation}
\delta N = N ~{\delta S \over S}=n ~\delta S,
\end{equation}

\noindent where $n$ is the number density. If we
assume that blend losses are predominantly due to close pairs of stars then the
mean number of close neighbours within $\delta S$ can be expressed as 
$\half n\delta S$. In other
words, we lose on average $\half n \delta S$ stars for every $\delta S$ 
area element
that has been found to contain a star. Consequently, for a detected set of
$N_d$ stars, the total blend loss will be

\begin{equation}
N-N_d=\half N_d~N~{\delta S \over S}
\end{equation}

\noindent Rearranging this and introducing $x=\delta S/S$ give the estimated total
number  of objects:

\begin{equation}
N={N_d \over 1-\half N_d~x}.
\end{equation}

\noindent Note that $\frac{1}{x}$ can be thought of as the number of resolution
elements in the image. 

\subsection{Monte Carlo simulations}

Equation~3 allows us to estimate the actual number of objects in our image,
based on the number we have detected and the size of our resolution elements.
We tested Eq.\ 3 with Monte Carlo simulations. Two different survey areas were
chosen to mimic real observations: 0.1 square degree with 1,000 objects and 1
square degree with 10,000 objects.  This way the surface density was kept 
constant and the effects of statistics could be checked. The confusion radius
was varied between 1$^{\prime\prime}$ and 40$^{\prime\prime}$. For each
simulation, we filled the survey area with $N$ randomly placed points
and retained only those, that fell outside the confusion area for all
previously placed points (i.e. after placing the first point, the second one
was kept only  if it was outside the confusion area of the first point; the
third one was kept  only if it was outside the confusion areas of the first two
points; and so on). Every simulation was repeated one hundred times and the
average blending losses were compared to those of calculated via Eq.\ 3.

\begin{figure}
\begin{center}
\includegraphics[width=80mm]{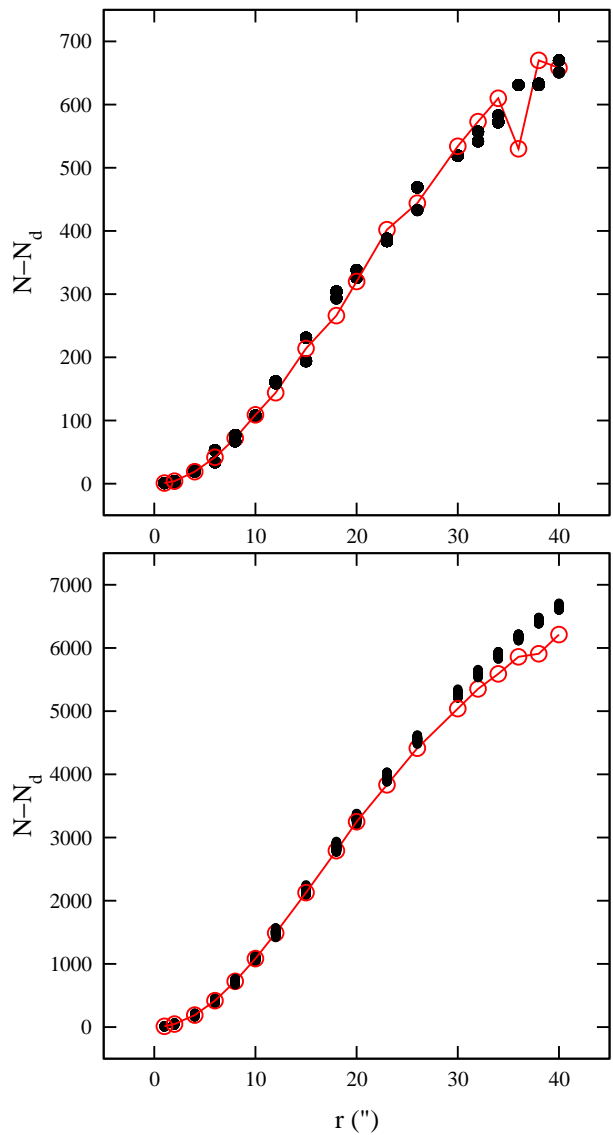}
\caption{Blend losses as a function of confusion radius in 0.1 deg$^2$ 
({\it top}) and 1 deg$^2$ ({\it bottom}) fields, where the 
stellar density was set to 10$^4$ star/deg$^2$ in both fields. The heavily
overlapping black dots represent one hundred simulations at each radius, 
while open circles connected with the solid lines show the predicted losses
from Eq.\ 3.}
\label{mcsim}
\end{center}
\end{figure}

The results are shown in Fig.\ \ref{mcsim}. Fractional losses range between 
0.1\% and over 65\%, with practically no difference between 1,000 and 10,000 
objects. The calculated blend losses are in excellent agreement with the true
values, except for the largest confusion radii. We stopped the simulations when
the integrated confusion area reached the full survey area, because after that
point, one can place infinitely many objects without detecting them. This is
why the loss tends to be underestimated after  $r=30^{\prime\prime}$. Also,
this is where the mean number of objects within a confusion area exceeds 2, so
that triplets and higher multiplets are no longer negligible. Keeping these
restrictions in mind, it is remarkable how well the blend losses agree with
Eq.\ 3. However, one question has to be addressed before we turn to real
datasets, for which binarity may be significant: to what extent can we assume
that seemingly random stellar fields can indeed be approximated by a uniform
random Poisson process? The answer can be found by checking the distance
distributions of all pairs in a real sample.

\subsection{The effects of double stars}

\begin{figure}
\begin{center}
\includegraphics[width=80mm]{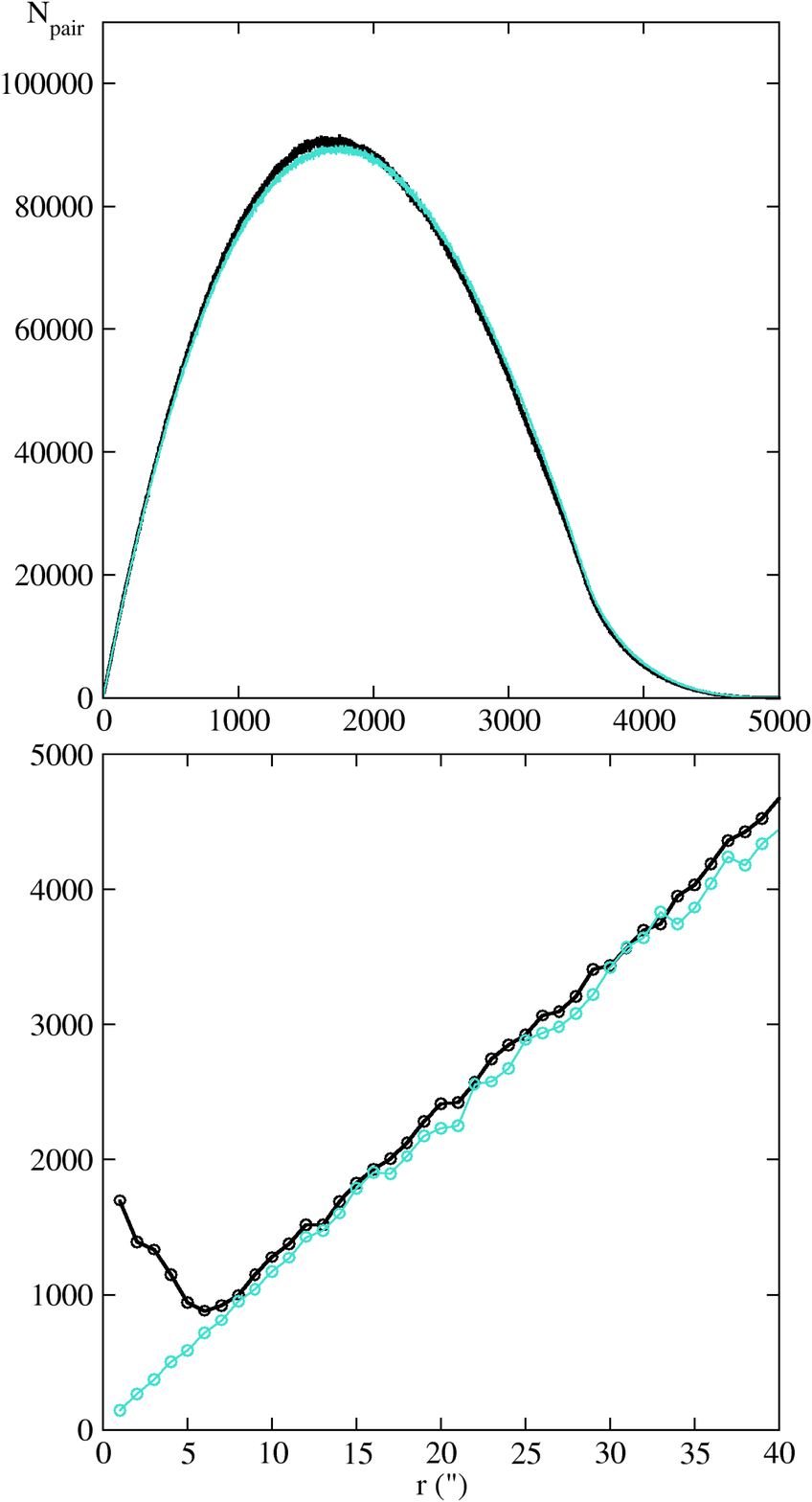}
\caption{Pair-separation distributions for a 1 deg$^2$ field around V705~Cas (black
line) and a random simulation (grey line). The lower panel is a zoom of the upper one
showing the presence of about 5,000 resolved double stars.}
\label{b2}
\end{center}
\end{figure}

We compare two histograms in Fig.\ \ref{b2}. The thick black line shows the
pair-distance distribution for USNO B1.0 stars around Nova~(V705)~Cas 1993 
(21,520 stars  with 14\fm00$<$red mag$<$17\fm99), while the grey line is for a
simulation with 21,600 random points with the same field of view. As expected,
the overall shapes of the distributions closely follow a
Poisson-distribution, with slight boundary effect for the large distances
(comparable to the diameter of the survey field), causing a somewhat sharper
cut-off than a pure Poissonian. 

We have three conclusions based on these kinds of comparisons for fields
in the Milky Way:

\begin{itemize}

\item[1.] The overall distributions agree very well with the pure random simulations.
There are slight indications for different shapes (see, e.g., the upper panel of
Fig.\ \ref{b2} around $r=1500^{\prime\prime}$ and $3000^{\prime\prime}$), 
but the differences hardly exceed the intrinsic scatter of the data.

\item[2.] The main disagreement occurs for the smallest pair-separations
(typically for $r<10^{\prime\prime}$), which must be due to a large number of
binary stars. For example, in the field around V705~Cas, the pair excess
suggests about 5000 visual double stars with separations less than
7$^{\prime\prime}$. In denser regions even higher numbers can occur: the 1
deg$^2$ USNO B1.0 field around Nova~(V4745)~Sgr~2003 contains about 20,000
double stars with red mags$<$17\fm0 and separations under
10$^{\prime\prime}$. Considering that the Washington Double Star Catalogue 
(Mason et al. 2001\footnote{A regularly updated version can  be found at\\ {\tt
http://ad.usno.navy.mil/wds/}}) contains approximately 100,000 astrometric
double and multiple stars, it is evident that most faint binary stars
have yet to be identified and catalogued (see also Nicholson 2002).

\item[3.] The contamination by physical binary stars can therefore
be a significant factor that may distort the results of random blending.

\end{itemize} 

\begin{figure}
\begin{center}
\includegraphics[width=80mm]{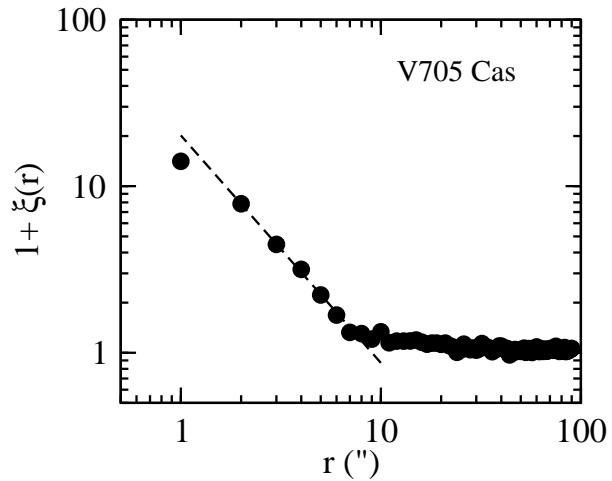}
\caption{The two-point correlation function for the 1 deg$^2$ field 
around V705~Cas. The dashed line shows a power-law fit to 
the data between 2 and 8 arcsecs ($1+\xi(r)\sim r^{-1.3}$).}
\label{2pcf}
\end{center}
\end{figure}

Star-star correlations can be taken into account with the two-point
correlation function, $\xi(r)$, which is defined by the joint probability of
finding an object in both of the surface elements $\delta S_1$ and $\delta S_2$ at
separation $r_{12}$:

\begin{equation}
P=n^2~\delta S_1~\delta S_2[1+\xi(r_{12})].
\end{equation}

\noindent Note that homogeneity and isotropy are assumed, so that $\xi$ is a function
of the separation alone (Peebles 1980). If we can calculate this excess
probability  for pairs, then multiplying random blend losses from Eq.\ 3 by
$(1+\xi(r_c))$ will correct the blending rate for the
double stars. Neglecting higher correlation functions is a reasonable
assumption, because the majority of stars belongs to single or double systems
(about 90\%, Abt 1983). Also, it is clear that even with this correction, the
calculated blending rate will be a lower limit, because we do not have any
information on the frequency of unresolved close binaries.

We can determine $\xi(r)$ using the recipe that leads to Eq.\ (47.14) in
Peebles (1980): place $N_t$ points at random in the survey area; let $n_p(t)$
be the number of pairs among these trial points at separation $r$ to 
$r+\delta r$, and let $n_p$ be the corresponding number of pairs in the real
catalogue of $N$ objects. Since $\xi=0$ for the trial points, the estimate of $\xi$ for
the data is:

\begin{equation}
1+\xi(r)={n_p \over n_p(t)} {N_t^2 \over N^2}.
\end{equation}

We show an example in Fig.\ \ref{2pcf}, which in our case is simply the ratio 
of the two histograms plotted in Fig.\ \ref{b2} (because $N_t$ was equal to
$N$). The statistical uncertainty of the first point at 
$r=1^{\prime\prime}$ is $\pm0.01$, far less than the symbol size, but 
a much larger
systematic error is likely because that radius is comparable to the image
scale of plates on which the catalogue is based. 
The dashed line in Fig.\ \ref{2pcf} shows a
power-law fit ($\xi(r)\sim r^\alpha$) for the linear regime of the 
log-log plot. This diagram suggests that for the given dataset of 21,500
stars, the  probability of finding two stars with 1 arcsec separation is about 20
times larger than for the pure random case. For a 1$^{\prime\prime}$ confusion
radius Eq.\ 3 gives $N-N_d=56$, which means the corrected blending loss,
including the measurable fraction of double stars, is about 1120 stars. In
other words, the chance of being a blend in this dataset is at least 5.2\%.

To conclude, it is always advisable to check star-star
correlations before applying Eq.\ 3, and only if $\xi(r)$ is close to zero
can we assume that double stars do not contribute significantly to 
blending (in practice, it means that the observations did not resolve 
binaries, probably because of the large distance).

\section{Magnitude bias in the OGLE-II observations of the Large Magellanic
Cloud}

Our first application is an analysis of the OGLE-II $BVI$ observations of the
Large Magellanic Cloud (Udalski et al. 2000). These data were obtained with 
the 1.3m Warsaw telescope at the Las Campanas Observatory for more than 7
million  stars in the central 4.5 deg$^2$ of the LMC. The completeness of the
resulting catalogue is high down to stars as faint as $B\approx20$ mag,
$V\approx20$ mag and $I\approx19.5$  mag. The median seeing of the observations
was 1\farcs3, with no observations made when the seeing exceeded 1\farcs6--1\farcs8
(Udalski et al. 2000). Here we estimate blend losses in typical fields in the
LMC and determine the corresponding magnitude biases in $V$ and $I$. 

This is a Case 1 blending, since no existing
catalogue has a smaller confusion radius than the OGLE-II
observations, except small field-of-view observations with the Hubble Space
Telescope (Olsen 1999), which were used to estimate  MACHO blend biases by Alcock
et al. (2001). To investigate possible photometric biases, we downloaded a
representative field from the OGLE public  archive\footnote{Available at {\tt
http://bulge.princeton.edu/$\sim$ogle/}}. We chose LMC\_SC5, centered at
RA(2000)=$5^{\rm h}23^{\rm m}48^{\rm s}$,
Dec(2000)=$-69^\circ41^\prime05^{\prime\prime}$, which contains about 460,000
stars.  The majority of stars are fainter than 18 mag (both in $V$ and $I$), so
we restricted the sample to stars between 18 mag and 21 mag. We used the
two-point correlation function to constrain possible binarity effects and
also to estimate the confusion radius.  Because the stellar density varies
gradually across  the bar of the LMC, we selected several subsamples
within which the density was constant. Here we
discuss two of them, one in   the middle of the bar (hereafter Region 1, $80.7<{\rm
RA[^\circ]}<80.8$,  $-69.7<{\rm Dec[^\circ]}<-69.6$) and one further south
(hereafter Region 2, $80.7<{\rm RA[^\circ]}<80.8$ and  $-70<{\rm
Dec[^\circ]}<-69.9$). Other subsamples yield practically identical  characteristics
for blending systematics.

\begin{figure}
\begin{center}
\includegraphics[width=80mm]{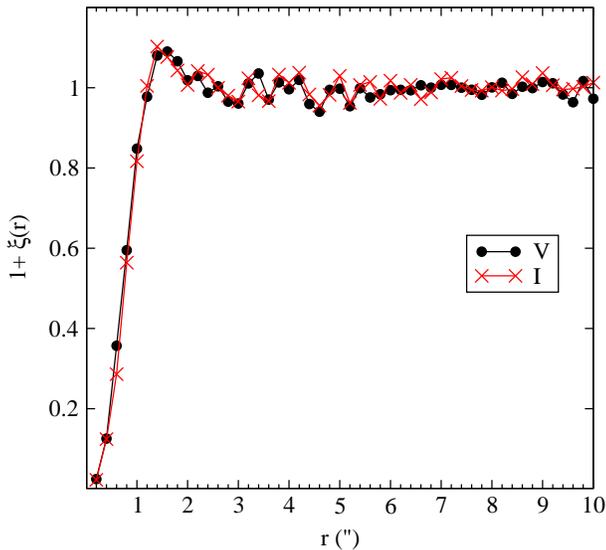}
\caption{Two-point correlation functions in $V$ and $I$ in Region 1.}
\label{ogle2pcf}
\end{center}
\end{figure}

\begin{table}
\begin{center}
\caption{Star counts, estimated total numbers of stars, blend losses and blending 
frequencies for two OGLE fields in the LMC.}
\label{ogleloss}
\begin{tabular}{cccccc}
\hline
Region & band & $N_d$ & $N$ & $N-N_d$ & $(N-N_d)/N_d$\\
\hline
1 & $V$ & 7598 & 9151 & 1553 & 20\% \\
2 & $V$ & 5354 & 6081 & 727 &  14\% \\
1 & $I$ & 7728 & 9340 & 1612 & 21\% \\
2 & $I$ & 6073 & 7026 & 953 & 16\% \\ 
\hline
\end{tabular}
\end{center}
\end{table}

In Fig.\ \ref{ogle2pcf} we show the two-point correlation functions   for
Region 1 in $V$ and $I$ bands. The step size in radius was $\delta r=0\farcs2$,
and we plotted the functions with linear scaling, so that changes around the
confusion radius could be noticed more easily. We do not see a significant
excess amount of correlated pairs, which is not surprising given the distance
to the LMC. There is a slight rise of the correlation function for $1\farcs4
\leq r \leq 3^{\prime\prime}$ that may be attributed to the widest 
binary stars. However, the correlation goes down quickly for $r<1\farcs0$.
Since the best seeing of the  OGLE images reached 0\farcs8 (Udalski et al.
1998), we adopt this value as the confusion radius. Our choice is also
consistent with ``critical radius'' chosen as 0\farcs75  by the OGLE-team,
within which objects were treated as identical (Udalski et al. 1998). 
The blend losses estimated using Eq.\ 3 are shown in Table\ \ref{ogleloss}.

\begin{figure*}
\begin{center}
\includegraphics[width=160mm]{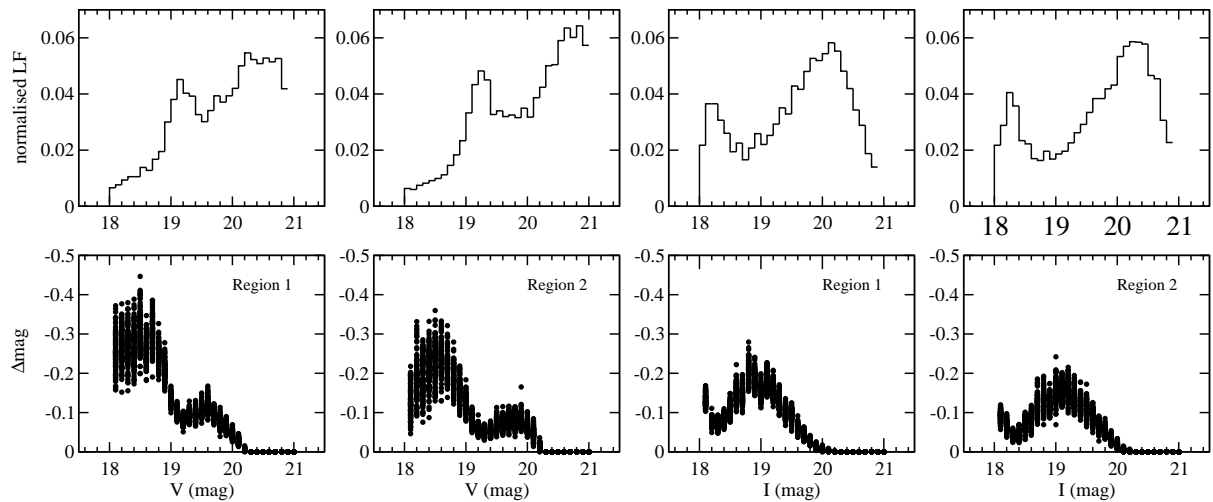}
\caption{The luminosity functions in Region 1 and 2 ({\it upper panels}) and the
calculated systematic magnitude biases ({\it lower panels}).}
\label{oglebias}
\end{center}
\end{figure*}

Apparently, about one in five-six stars is a blend having an unresolved
companion that is within 3 mags in brightness (since our data sample covers the
range 18 to 21 mag). We have to stress that these numbers are only lower limits
because of the incompleteness of the OGLE data below $\sim$20 mag, which means
the systematic errors we derive are still underestimated. Our main
purpose is to point out the existence of surprisingly large systematic errors
that can be determined  from a single dataset alone. Deeper data (like in
Alcock et al. 2001) would  only shift these systematics toward larger values,
but that is beyond the scope  of the present paper.

Having estimated the fraction of blended stars, we are now in a position to
calculate the extent to which they will bias the photometry. We assume that the
probability of a star blending with a neighbour does not depend on its
brightness, so that blended and unblended stars have the same luminosity
function (LF) within each region. This is likely to introduce a slight
systematic error, but we estimate this to have a marginal effect on the final
outcome of the investigation.  Upper panels in Fig.\ \ref{oglebias} contain the
normalized $V$ and $I$ luminosity functions ($\sum_{18}^{21}{\rm LF}(m)=1$),
where the prominent peaks at $V\approx19$ mag and $I\approx18$ mag correspond
to the red clump (Udalski 2000).  The $I$-band decline for $I>20$ mag shows the
decreasing completeness (see Table\ 4 in Udalski et al. 2000), which is 
less severe in $V$.  The next step was to carry out a Monte Carlo
simulation for each region and each filter: we placed $N$ stars at random into the given survey area, where $N$
is the corresponding number in the fourth column of Table\ \ref{ogleloss}.
The luminosity function was set to the one shown in Fig.\ \ref{oglebias}.
Then we determined which stars were ``blends'' in the random set and calculated
the integrated magnitude of each blend. The magnitude difference between the
blend and the brightest star within it was assigned as a
bias at the integrated magnitude. Finally, these bias values were averaged for
every 0.1 mag  bin of the luminosity function (unblended stars were  taken into
account by adding zero bias). The whole procedure was repeated one hundred
times and the  results are shown in the lower panels of Fig.\ \ref{oglebias}.
Small changes to the input luminosity functions affected the final bias
distributions at a level lower than the scatter visible in the plots. 

We can draw several conclusions based on Fig.\ \ref{oglebias}. Firstly and most
importantly, blending introduces systematic errors as large as 0\fm2--0\fm3
(for instance, the mean bias between $V=18-19$ mag is 0\fm25 in Region 1 and
0\fm18 in Region 2), which can have serious consequences if neglected. Details
of OGLE-II reduction, tests of photometric accuracy and incompleteness were
published in Udalski et al. (1998) and they did not mention random blending as
a possible source of systematic biases.  Very recently, Alcock et al. (2004)
presented an analysis of first-overtone RR~Lyrae stars in the MACHO database,
for which systematic magnitude biases due to blending were also estimated. At
$V\approx19\fm3$,  where RR~Lyrae stars concentrate, Alcock et al. (2004)
arrived at $\Delta V$ ranging from $-0\fm11$ to $-0\fm21$ for various
assumptions (they made artificial star tests for various stellar densities and
calculated the differences between the input and recovered magnitudes of
artificial stars).  Our results are based on a very different approach applied
to a very different dataset, and the agreement suggests that random blending 
must be taken into account in such high-level crowding as the present one.
Secondly, the gradual decrease of the bias towards fainter magnitudes shows the
effect of incompleteness: as was shown by Alcock et al. (2001) for the MACHO
database, systematic errors increase monotonically over the examined magnitude
range. For that reason, the calculated biases between 18 and 19 mag can be
considered as lower limits for the fainter magnitudes; the magnitude range of
RR~Lyrae stars is therefore very likely to suffer from a bias up to 0\fm2--0\fm3,
which has been neglected in the past. This leads to our third conclusion:
results based on the OGLE-II data that supported the LMC ``short'' distance
scale ($(m-M)_{\rm LMC}\approx18\fm3$) are likely to suffer from this
systematic error.  For example, Udalski (2000) calibrated the zero-point of the
distance scale using  RR~Lyrae variables and red clump stars to be 
$(m-M)_{\rm LMC}=18\fm24$ with 0\fm07 systematic error. However, blending
correction significantly decreases the apparent conflict between this and the ``standard''
value  (18\fm5, e.g. Alves 2004), which shows the importance of blending biases.

It must be stressed that ground-based observations of more distant galaxies may
suffer from  even larger systematic errors. Detecting variable stars by the
image subtraction method circumvents the problem (see a recent application by 
Bonanos \& Stanek 2003), but measuring apparent magnitudes (and hence distance)
requires use of  space telescopes. Calculations like the present one can help
estimate systematics but more reliable results for nearby galaxies need
high-resolution observations.

\section{Field star distributions around seven recent novae}

\begin{table}
\begin{center}
\caption{Seven bright nova outbursts with reported progenitor candidates in the 
last decade.}
\label{novae}
\begin{tabular}{llc}
\hline
Star (year) & Progenitor details & Ref.\\
\hline
V1494~Aql (1999/2) & Red mag 15\fm6 & 1 \\
V2275~Cyg (2001) & Red mag 18\fm8 (USNO A2.0) & 2 \\
V382~Vel (1999) & Quiescent magnitude $V=16\fm56$ & 3\\
V4743~Sgr (2002/2) & Red mag 16\fm7 (USNO A2.0) & 4\\
V4745~Sgr (2003) & DSS red plate, $R\approx17\fm9$, apparently &\\
                 & blended with a faint companion to SE & 5 \\ 
V705~Cas (1993) & Candidate precursor of red mag about& \\
               & 17\fm0, northern component of a merged & \\ 
	       & pair with separation about 2 arcsec & 6, 7 \\ 		 
V723~Cas (1995) & DSS red plate, mag 18-19 & 8\\
\hline
\end{tabular}
\end{center}
References: 1 -- Pereira et al. (1999); 2 -- Schmeer (2001); 3 -- Platais et al. (2000);
4 -- Haseda et al. (2002); 5 -- Brown et al. (2003); 6 -- Skiff et al. (1993);
7 -- Munari et al. (1994); 8 -- Hirosawa et al. (1995)
\end{table}

\begin{figure*}
\begin{center}
\includegraphics[width=140mm]{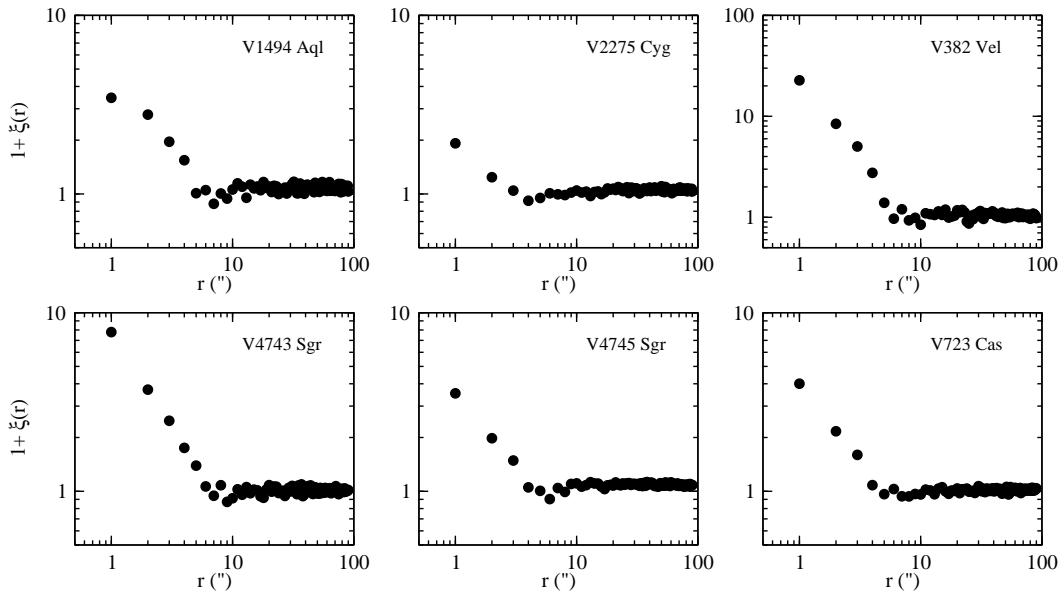}
\caption{Two-point correlation functions around six novae.}
\label{2pcfs}
\end{center}
\end{figure*}

\begin{table*}
\begin{center}
\caption{Statistical properties of the examined datasets. Abbreviations: $N_d$ -- number of stars
within the specified magnitude range; $N(m<m_u)$ -- total number of stars brighter than the
fainter limit; $\Delta N$ -- random blend loss via Eq. 3; $\xi(r_c)$ -- the two-point
correlation function at $r_c=1\farcs5$; blend rate -- blending occurrence rate, corrected
by $1+\xi(r)$; A1 --
the probability that the pre-outburst star coincides with the nova by chance; A2 -- the probability that
the optical companion is correlated with the nova itself.}
\label{novaresult}
\begin{tabular}{llrrlcrrc}
\hline
Star & Mag range & $N_d$ & $N(m<m_u)$ & $\Delta N$ & $1+\xi(r_c)$ & blend & A1 & A2\\
     &           &       &            &            &              & rate  &    &    \\
\hline
V1494~Aql & 15--16 & 8,333 & 13,229 & 19 & 4 & 0.9\% & 0.5\% & 75\%\\
V2275~Cyg & 18--19 & 22,249 & 47,255 & 137 & 1.5 & 0.9\% & 1.2\% & --\\
V382~Vel & 16--16.6 & 6,125 & 19,291 & 10 & 15 & 2.4\% & 0.3\% & --\\
V4743~Sgr & 16--17 & 10,775 & 20,165 & 32 & 5 & 1.5\% & 0.6\% & --\\
V4745~Sgr & 17--18.5 & 35,997 & 62,042 & 485 & 3 & 4\% & 2.7\% & -- \\
V705~Cas & 16.5--17.5 & 8,066 & 17,374 & 18 & 10 & 2.2\% & 0.5\% & 90\%\\
V723~Cas & 18--19.5 & 18,932 & 31,302 & 99 & 3 & 1.6\% & 1\% & --\\ 
\hline
\end{tabular}
\end{center}
\end{table*}

Our next case study was inspired 
by the independent discovery of the close optical companion of nova 
V1494~Aql (Kiss et al. 2004). We selected several bright nova outbursts in the last decade
for which  progenitor star candidates have been identified in Digitized Sky Survey
plates or deep star catalogues (most often in the USNO A1.0 and A2.0 catalogue
releases). The available details on these candidates are summarized in Table\
\ref{novae}.  We examined the following questions: How probable is it that the
candidates are unrelated stars located only by chance at the nova coordinates?
In cases where a star was found a few arcseconds from the nova position, 
(like for V1494~Aql and V705~Cas), how probable
is is that they are physically related to the nova system? 

To find out the answers, we downloaded and analysed 1 deg$^2$ USNO B1.0 fields 
around each nova using the USNO Flagstaff Station Integrated Image and
Catalogue Archive Service \footnote{\tt http://www.nofs.navy.mil/data/fchpix/}.
These fields provide a complete coverage down to $V=21\fm0$, 0\farcs2
astrometric accuracy at J2000, 0.3 mag photometric accuracy in up to five
colours and 85\% accuracy for distinguishing stars from nonstellar objects (see
Monet et al. 2003 for more details). The latter issue is less relevant in these
fields because of the low galactic latitudes. In two cases we found
incompleteness of the data, either in the form of large empty areas in the 
given field (V4745~Sgr) or sudden jumps in the stellar density within certain
rectangular areas (V4743~Sgr). Both fields are located in the densest regions of
the  Milky Way, which explains the difficulties of measuring wide-field
Schmidt-plates in such crowded fields. We took these into account in our
analysis (e.g. in the latter case we omitted patches of lower
density, yielding a decreased survey area). 

Since this is Case 1 blending, we calculated random blend losses for each nova by 
choosing all stars within 1--1.5 mag in brightness of the reported progenitor
(when the accuracy was worse, we chose the wider range).  For this, we
used red magnitude data in the catalogue. We assumed a 1\farcs5 confusion radius,
based on the  pixel scale of the scanned Schmidt-plates (usually about
0\farcs9/pixel,  Monet et al. 2003) and the fact that the optical companion of
V1494~Aql located at 1\farcs4 was unresolved in the data. Two-point correlation
functions (Fig.\ \ref{2pcfs}) were determined via Eq.\ 5, filling up the survey areas by 25,000 points
at random. Blend losses were then multiplied by the two-point correlation functions
at $r=1\farcs5$. Finally, we calculated the chance of finding a random blend within 1\farcs5 of
the nova coordinates. We summarize the results in
Table\ \ref{novaresult} and in  Fig.\ \ref{2pcfs} (for V705~Cas we got essentially
the same as in  Fig.\ \ref{2pcf} for a wider magnitude range, so that there was no
reason to repeat the plot).

It is evident from Fig.\ \ref{2pcfs} that the samples are dominated by binary
stars in every field for separations under $\sim10^{\prime\prime}$. The
highest correlation was found around V382~Vel, where the number of 1-arcsec
doubles exceeds the number expected for a random field by a factor of 23! In other
words, at least 22 of every 23 1-arcsec pairs are physically related (binary)
stars. It is also prominent that there is a slope decrease in every correlation
function for the smallest radii, which we interpret as a result of confusion
losses. For that reason, we estimated $\xi(r_c=1\farcs5)$ (6th column in Table\
\ref{novaresult}) from extrapolated linear fits over the log-log plot of the 2--6
arcsecs data, rather than from interpolations between 1 and 2 arcsecs.

The numbers in Table\ \ref{novaresult} clearly show that random blend loss 
stays well below 1\%; even for the worst case, V4745~Sgr, it is only 1.4\% of
$N_d$.  After multiplying by the estimated two-point correlation function
values, the blending rate still remains  around 1--2\%. The eighth column in
Table\ \ref{novaresult} (A1) confirms that progenitors can be identified with
high confidence, even in the central regions of the Milky Way. 

\begin{figure*}
\begin{center}
\includegraphics[width=140mm]{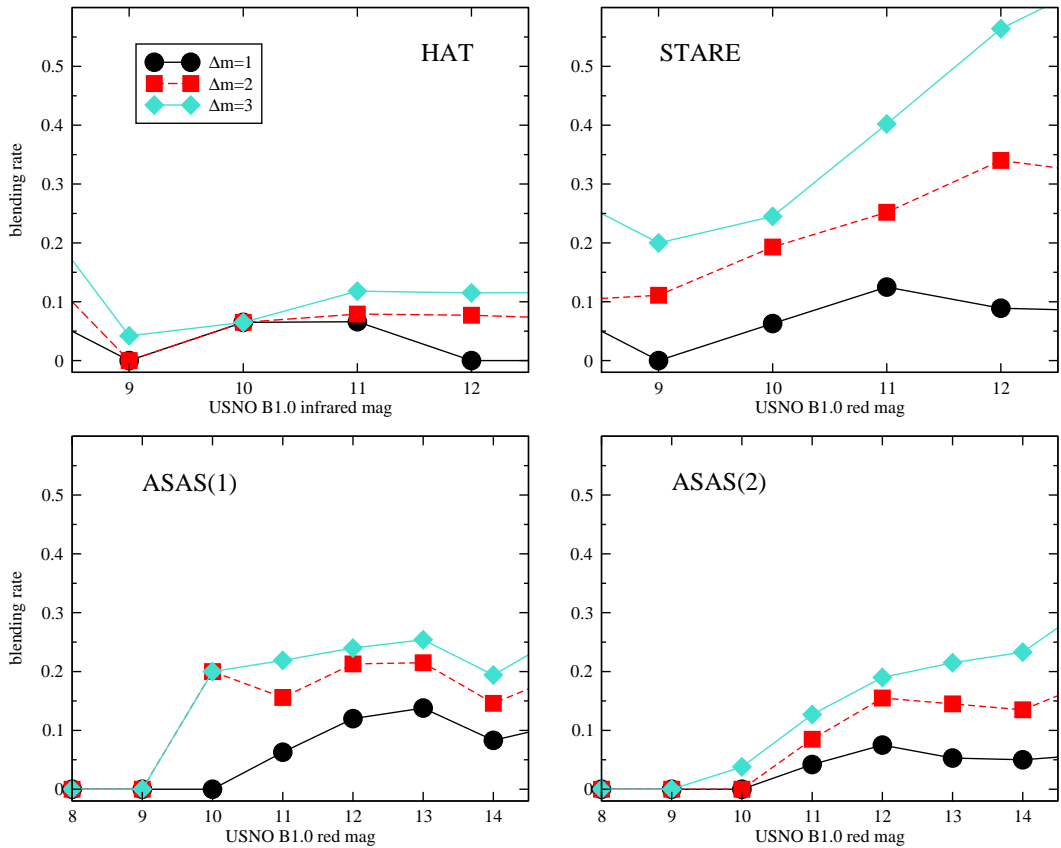}
\caption{Blending rates for similar projects at four different galactic latitudes
(HAT: $+30^\circ$; STARE: $+2^\circ$; ASAS(1): $-60^\circ$; ASAS(2): $-20^\circ$). The
confusion radius is about $20^{\prime\prime}$ for each project.}
\label{surveys}
\end{center}
\end{figure*}

The most interesting numbers can be found in the last column of  Table\
\ref{novaresult}. They follow from the probability interpretation of the 
two-point correlation function. Since $(1+\xi(r))$ reflects the excess probability
for finding  a pair in a sample at distance $r$, $1+\xi(r)=4$ (first row in Table\
\ref{novaresult}) means that there are four times more pairs than in a random
sample, so that for any given pair the chance of being correlated is 75\%.
Therefore, these data suggest that it is quite possible  that the optical
companions of V1494~Aql and V705~Cas form wide hierarchical triple systems with
the cataclysmic components. Interestingly, for V1494~Aql this is also  supported
by our knowledge of the optical companion. In Kiss et al. (2004) we assigned  late
F-early G as an approximate spectral type of the companion. The corresponding
absolute magnitude for a main sequence star is about $M_V=$+4\fm0, which agrees
within the error bars  with the quiescent absolute magnitude of the nova (Kiss et
al. 2004). Hence, the close apparent magnitudes of the pre-nova and the companion
suggest similar distances, consequently the possibility of physical correlation.
For V705~Cas, we do not have similar supporting evidence, but it is an intriguing
possibility that these optical components  are in triple systems with the novae.
If confirmed, their presence could be used,  for example, to derive independent
distances via spectroscopic parallaxes, which in turn would allow one to calculate
accurate absolute magnitudes of the nova eruptions. 

This case study shows that seeing-limited images are not significantly affected
by random blending, even in the galactic plane. We now turn to the case of
images with much lower spatial resolution.

\section{Wide-field photometric surveys}

In the recent years there has been an explosion in the number of small and
medium-sized robotic telescopes  that monitor selected regions of the sky with
various instruments (e.g., Hessman (2004) listed 80 different
projects). We selected three  representative small
instruments that are currently running, to investigate Case 2 blending. They
are:

\begin{enumerate}

\item The Hungarian Automated Telescope (HAT) which, in its present status, observes stars
between  $I=9-12$ mag in selected fields of the northern sky. The data were reduced with
PSF-fitting photometry (Bakos et al. 2004). We
downloaded the 1 deg$^2$ USNO B1.0 field in Hercules (RA(2000)=$17^{\rm h}36^{\rm m}$,
Dec(2000)=$37^\circ30^\prime$, $b\approx30^\circ$). 

\item The STellar Astrophysics \& Research on Exoplanets (STARE) project, observing between
$R=9-12\fm5$ in various fields (Alonso et al. 2003, Brown
2003). We downloaded the 1 deg$^2$ USNO B1.0 field, centered at RA(2000)=$20^{\rm h}06^{\rm
m}$, Dec(2000)=$36^\circ00^\prime$ ($b\approx2^\circ$). 

\item The All Sky Automated Survey (ASAS), which covers over 1 million stars in the
southern hemisphere between $V=8\fm5-15$ mag. The data are reduced with simple aperture
photometry (Pojmanski 2002, 2004). We selected  two 1-deg$^2$ fields at random at two
different galactic latitudes: RA(2000)=$00^{\rm h}19^{\rm m}$,
Dec(2000)=$-56^\circ00^\prime$ ($b\approx-60^\circ$) and  RA(2000)=$06^{\rm h}10^{\rm m}$,
Dec(2000)=$-23^\circ00^\prime$ ($b\approx-20^\circ$).

\end{enumerate}  

Each project is characterized by $r_c\approx20^{\prime\prime}$, so that their
comparison reveals the differences that depend on the galactic latitude. To
match the photometric bands used in these projects, we examined USNO B1.0
infrared magnitudes for the HAT-field and red magnitudes for the remaining
regions. The main aim was to get an idea of the fraction of biased stars,
because although crowding problems are well-known for these instrumental
setups, we did not find any quantitative description of the issue. Therefore,
we calculated blending rates for various $m$ and $\Delta m$ values that covered
the magnitude ranges of the projects. This rate was defined as the following
ratio:

\begin{equation}
{\rm blending~rate}={\nu(m,\Delta m,r_c) \over \nu(m)},
\end{equation}

\noindent where $\nu(m,\Delta m, r_c)$ is the number of stars which have
fainter neighbours within a distance $r_c$ and magnitude difference $\Delta m$;
$\nu(m)$ is the total number of $m$ magnitude stars (in our case, defined as
stars with apparent brightnesses within $m$ and $m+1$ mag).  We assumed
negligible blend losses in the USNO B1.0 catalogue over the studied magnitude
and separation ranges. The results are shown in Fig.\ \ref{surveys}. We also
determined two-point correlation functions, which showed that for these bright
magnitudes the overwhelming majority of pairs closer than the confusion radius
are physically related double stars. 

Figure \ref{surveys} implies an alarming rate of blending, even for high 
galactic latitudes. We see that 10 to 20 percent of objects observed by the HAT
and ASAS projects have blends within 3 mags, while up to 50\% are affected
near the galactic plane (STARE). Correlated pairs can make the situation
quite bad even for the brightest stars, which means blending must be taken into
account in every case. Presently available star catalogues offer a good
opportunity to do that, thus we recommend to add blending information in all
cases when finally reduced data are made accessible to the wider community.
Also, the implementation of the image subtraction method (Alard \& Lupton 1998)
is highly desirable in this type of project, because  the photocenter of
the variable source can be used to identify which object within a blend is 
varying (Alard 1996; see also Hartman et al. 2004). 

\section{Conclusions}

In this study we investigated the effects of random blending that involves
stars of similar brightnesses, leading to significant biases in the measured
apparent magnitudes and  amplitudes. A simple formula (Eq.\ 3) was derived to
calculate blend losses in a real catalogue of stars based on the confusion
radius, total number of detected stars and survey area. We showed that the
two-point correlations must be included in calculations when studying galactic
fields, where binary stars dominate for separations under 10$^{\prime\prime}$.
Outside the Milky Way, we quickly lose the information on wide binaries, so
that the pure random case applies. That is why the calculated blending rate
always puts only a lower limit to the full blending. 

We discussed three different applications, which
demonstrate  the importance of these phenomena. The most interesting
results were presented  for the OGLE-II data of the Large Magellanic Cloud.
With extremely high stellar angular densities, these observations are much more
biased by random blending than those fields in the galactic plane. Even though
no observations were carried out for seeing larger than 1\farcs6--1\farcs8, we
estimated that 15-20\% of the sample are affected by blends that are within three
magnitudes in apparent brightness. The resulting magnitude biases can reach
0\fm2--0\fm3, depending on the angular position within the
LMC. This allows us to reconcile the OGLE-based
distance moduli of the LMC ($\mu\approx18\fm3$) and the generally
adopted ``standard'' one ($\mu\approx18\fm5$). Random blend calculations are
therefore highly desirable in every case when significant blending is expected,
especially in extragalactic observations.

The analysis of star fields around seven novae showed that progenitor
identification is very secure, even in the extremely dense fields of the Bulge.
Furthermore, statistical evidence points toward the existence of wide
hierarchical triple systems, in which the third 
component lies at 1--2$^{\prime\prime}$ (i.e. several thousand AU)
from the novae. Thus, it is highly desirable to measure proper
motions for V1494~Aql and V705~Cas and their optical companions to confirm or
reject the physical correlations.

In our last example we demonstrated  high-blending rates in typical wide-field
photometric surveys for exoplanet transits. In certain cases, up to 50--60\% of
stars can be affected by blending objects within 3 magnitudes. We recommend that
cross-correlation with appropriate star catalogues should always be done as
part of the regular data reduction, to flag every possibly problematic star.

Our presented method for calculating blend losses is very simple, thus may not be
applicable in certain cases. For instance, when the stellar angular density shows
strong gradient, like in a globular cluster or outer parts of a resolved galaxy,
the basic assumption of homogeneity and isotropy will be invalid. A possible
solution in such cases is to introduce a position-dependent density $n=\rho(x,y)$,
where $x$ and $y$ are the image coordinates,  and to integrate all equations over
the inhomogeneous and anisotropic sample. In principle, the numeric implementation
of this generalisation is not too difficult. Another way to treat these cases 
is
to split the data into small segments within which $n\approx{\rm const.}$ and, 
after calculating blending biases, take averages over the segments. 

To summarize, the experience with real datasets suggests that it is always  highly
advisable to estimate random blend losses when the star density and/or the confusion 
radius have relatively large values -- Eq.\ 3 or deep star catalogues can be used to 
characterize any specific example. Where applicable, 
the upgrade from aperture and PSF-photometry to the image subtraction method is 
desirable to reduce some of the blending effects.

\section*{Acknowledgments} 

This work has been supported by the FKFP Grant
0010/2001, OTKA Grant \#T042509 and the Australian Research Council. 
Thanks are due to Dr. A. Udalski, whose comments helped improve the paper.
LLK is supported by a University of Sydney Postdoctoral Research Fellowship.  
The NASA ADS Abstract Service was used to access data and references.

\end{document}